\documentclass[a4paper,10pt]{article}
\usepackage[dvips]{graphicx}
\usepackage{a4wide}
\usepackage{bbm}
\usepackage{epsfig} 
\usepackage{amsmath}
\usepackage{color}

\definecolor{blue}{rgb}{0,0,1}

\title{Pair production in inhomogeneous fields}
\author{Holger Gies and Klaus Klingm\"uller \\ \\
     \small \it Institut f\"ur theoretische Physik, Universit\"at Heidelberg
     \\ \small \it Philosophenweg 16, D-69120 Heidelberg, Germany 
}

\begin{document}

\maketitle

$\text{}$

\vspace{-8.7cm}

{\hfill \small\sf HD-THEP-05-07, 
{  }http://arXiv.org/abs/hep-ph/0505099} 

\vspace{7.8cm}

\begin{abstract}
  We employ the recently developed worldline numerics, which combines
  string-inspired field theory methods with Monte Carlo techniques, to
  develop an algorithm for the computation of pair-production rates in
  scalar QED for inhomogeneous background fields.  We test the
  algorithm with the classic Sauter potential, for which we compute
  the local production rate for the first time.  Furthermore, we study
  the production rate for a superposition of a constant $E$ field and
  a spatially oscillating field for various oscillation frequencies.
  Our results reveal that the approximation by a {\em local}
  derivative expansion already fails for frequencies small compared to
  the electron mass scale, whereas for strongly oscillating fields a
  derivative expansion for the {\em averaged} field represents an
  acceptable approximation.  The worldline picture makes the nonlocal
  nature of pair production transparent and facilitates a profound
  understanding of this important quantum phenomenon.
\end{abstract}

\section{Introduction}
Pair production was first proposed for electron-positron pairs in strong,
temporally and spatially constant electric fields
\cite{Sauter:1931,Heisenberg:1935qt,Weisskopf:1936}. Today it is often
referred to as the Schwinger \cite{Schwinger:1951nm} mechanism.  As a
nonperturbative mechanism, pair production is of great theoretical interest.
From a phenomenological point of view, it corresponds to probing the theory in
the domain of strong fields.  Consequently, we encounter pair production in
many topics of contemporary physics, for instance, black hole evaporation
\cite{Hawking:1974rv} and $e^+e^-$ creation in the vicinity of charged black
holes \cite{Damour:1974qv,Kim:2004us} as well as particle production in
hadronic collisions \cite{Casher:1978wy} and in the early universe
\cite{Parker:1969au,Bjoern}.  Since QED pair production in strong fields
represents the conceptually simplest case, it can serve as a theoretical
laboratory for all these cases.

A sizeable rate for spontaneous pair production requires extraordinary strong
electric fields, comparable in size to the so-called critical field strength,
which corresponds to the electron-mass scale, $E_\mathrm{cr}=m^2/e\approx
1.3\cdot10^{18}{V\over m}$.  For a long time, it seemed inconceivable to
produce macroscopic electric fields of the required strength in the
laboratory, but today, with the development of strong lasers, there are
several promising experiments in progress
\cite{Arthur:1998yq,Materlik:2001qr,Brinkmann:2002bt}; for a discussion of
experimental requirements, see \cite{Ringwald:2003iv}.

Many different theoretical methods, such as the propertime method
\cite{Schwinger:1951nm,DeWitt:1975ys}, WKB techniques
\cite{Brezin:1970xf,Popov:1972,Popov:1973az,Piazza:2004sv}, the
Schr\"odinger-Functional approach \cite{Hallin:1994ad}, functional
techniques \cite{Fried:2001ur,Avan:2002dn}, kinetic equations
\cite{Smolyansky:1997fc,Smolyansky:1997ji,Kluger:1998bm,Alkofer:2001ik},
various instanton techniques
\cite{Affleck:1981bm,Kim:2000un,Kim:2003qp,DunneSchubertInPrep}, Borel
summation \cite{Dunne:1999uy,Dunne:2002qg}, and propagator
constructions \cite{Dietrich:2003qf,Dietrich:2004eb}, have been
developed to study pair production in external fields. Also,
finite-temperature contributions have been determined which first
occur at the two-loop level \cite{Gies:1999vb,Dittrich:2000zu}. Of
particular conceptual interest is the production rate in terms of the
effective action for a given background, which is also used in this
work.  Owing to an intimate relation between the effective action and
the vacuum-persistence amplitude, it is the imaginary part of the
effective action that encodes information about pair production which,
in this context, is interpreted as spontaneous vacuum decay.  This
approach yields the instantaneous production rate, neglecting
back-reactions and memory effects. However, this rate can serve as a
source term for kinetic equations, which can then take back-reactions
and memory effects into account
\cite{Smolyansky:1997fc,Smolyansky:1997ji,Kluger:1998bm,Alkofer:2001ik}.

Even though the existing methods follow a well defined and technically
stringent concept, their application often faces serious technical and
conceptual difficulties.
Up to now, no reliable and universal method---be it analytic or numeric---is
available for the calculation of pair-production rates in inhomogeneous
electric fields.
In standard approaches, functional traces have
to be evaluated with the knowledge of the spectrum of the corresponding
differential operator, which is only available for special cases. Moreover,
controlling the divergencies that possibly occur upon summing up the
eigenvalues is a delicate task.

In the present work, we solve these problems by using the recently
developed numerical worldline techniques
\cite{Gies:2001zp,Gies:2001tj,Langfeld:2002vy,Gies:2003cv,Moyaerts:2003ts}
which are based on the string-inspired worldline formalism
\cite{Feynman:1950ir,Halpern:1977he,Polyakov:1987ez,Bern:1991aq,Strassler:1992zr,
Schmidt:1993rk,Schmidt:1994aq,Reuter:1996zm,Schubert:2001he}.  The
important advantage compared to other approaches lies in the fact that
worldline numerics can be formulated independently of any symmetry of
the background. The identification of and the summation over the
spectrum of quantum fluctuations are done in one single and finite
step. For simplicity, we confine ourselves to scalar QED;
generalization to spinor QED is, in principle, straightforward and
will be discussed below.

Beyond the computational advantages of worldline techniques, the worldline
picture also helps to understand conceptual aspects in more depth. In
particular, the nature and the role of nonlocalities become highly transparent
from the worldline viewpoint, since the worldlines themselves represent
extended virtual trajectories of the fluctuating particles in coordinate
space. In the present context, we are aiming at the quantum effective action
which, of course, receives nonlocal contributions in general. However, many
standard approximation methods suppress (or shade) nonlocalities by
construction, as, e.g., the derivative expansion. Hence, pair production as
described by the Schwinger formula is often recognized as a nonperturbative
phenomenon, but not so much as a nonlocal phenomenon. Nevertheless, the latter
property is crucial, as the following heuristic argument elucidates: in order
for a virtual pair to become real, i.e., on-shell, the pair must gain at least
the amount of $2m$ of energy; this is only possible by propagating in opposite
directions in the electric field. This delocalization of the pair wave
function is mandatory for gaining sufficient energy.

In constant electric fields, this delocalization remains invisible in the
final result. By contrast, in inhomogeneous fields the spacetime dependence of
the delocalized wave function matters a great deal and can even dominate the
resulting effect, as our results demonstrate. In the worldline picture, the
nonlocal effects already become transparent on the level of the formalism,
since the extended worldlines exactly describe the delocalization of a virtual
pair. 

At this point, we would like to stress the difference of the present
work to earlier applications of worldline numerics. Whereas the
algorithms developed so far in
\cite{Gies:2001zp,Gies:2001tj,Langfeld:2002vy,Gies:2003cv,Moyaerts:2003ts}
have proven their capabilities for computing the real part of the
effective action (and action densities), the computation of the
imaginary part is by no means a straightforward generalization. The
reason for this lies in the truly Minkowskian nature of the problem of
pair production: vacuum decay only occurs for real, i.e., Minkowskian,
electric fields. This contrasts with the indispensable necessity of a
Euclidean formulation for solving the worldline integrals by a
statistical Monte Carlo algorithm. In practice, this results in an
overlap problem: the finite Euclidean worldline ensemble can have
little overlap with those worldlines that contribute dominantly to
Minkowski-valued observables. We solve this fundamental problem by
resorting to a technique developed in
\cite{GiesSanchez-GuillenVazquez_in_preparation} in the different
context of nonperturbative Euclidean worldline numerics: we fit a
suitable cumulative density function (CDF) of the Euclidean ensemble
to a physically motivated ansatz that can be continued analytically to
Minkowski space. We should emphasize that this continuation represents
an extrapolation of certain ensemble properties to Minkowski space
which is an a priori uncontrolled procedure resulting in systematic
errors. We check this extrapolation carefully against various
analytically known results and find negligibly small systematic errors
compared to the statistical Monte Carlo errors. Hence, we regard the
overlap problem as solved for the present problem. This solution is
obtained at the expense of numerical cost; moreover, the algorithm
can, in principle, not be made arbitrarily precise, in contrast to
former applications of worldline numerics. Nevertheless, for the
problem of pair production and as far as the experimentally required
accuracy is concerned, we believe that our algorithm is sufficiently
powerful.

\section{Worldline formalism for pair production}
The vacuum-persistence amplitude can be related to the effective
action $\Gamma_{\text{M}}$ in Minkowski space,
\[\langle\Omega|e^{-iHT}|\Omega\rangle=e^{i\Gamma_\mathrm{M}}.\]
The corresponding probability for the vacuum to decay spontaneously is
\[P=1-e^{-2\mathrm{Im}\Gamma_\mathrm{M}}.\]
In the case of QED with electric background fields, vacuum decay
occurs in the form of spontaneous pair production, the production rate
per unit time and volume of which is directly proportional to the
imaginary part of the effective action density (effective Lagrangian).

In scalar QED, the one-loop contribution to the
Euclidean effective action $\Gamma_{\text{E}}$ reads
\begin{equation}
  \label{eq:Hw}
  \Gamma_\mathrm{E}^1[A]=\ln\mathrm{det}\left(-(\partial+ieA)^2+m^2\right),  
\end{equation}
where  $\Gamma_{\text{M}}$ and $\Gamma_{\text{E}}$ differ by a
minus sign, $\Gamma_{\text{M}}=-\Gamma_{\text{E}}$. In the worldline
approach, the logarithm of the determinant in $D$-dimensional
spacetime is represented by a path integral \cite{Schubert:2001he},
\begin{equation}
  \Gamma_\mathrm{E}^1[A]=- \frac{1}{(4\pi)^{D/2}}\int_0^\infty{dT\over
T^{1+D/2}}\ e^{-m^2T} \int_{x(0)=x(T)}\mathcal
Dx(\tau)\ e^{-\int_0^T d\tau\ \left({\dot x^2\over4}+ie\dot
xA(x)\right)}\label{ga},
\end{equation}
where the integration parameter $T$ is called the propertime.  The
path integral runs over all closed worldlines, parameterized by the
propertime. The worldlines can be viewed as the trajectories of the
virtual fluctuations in coordinate space.  The path integral is
normalized to give 1 in the limit of zero gauge potential.  We split
the path integral into an integral over all paths with a common center
of mass $x_0$ and an ordinary integral over all $x_0$,
$x(\tau)\rightarrow x_0+x(\tau)$, where $\int_0^Td\tau\ x(\tau)=0$.
Introducing the \emph{Wilson loop},
\begin{equation}
  W_{x_0}[x(\tau)]:=e^{-ie\int_0^Td\tau\ \dot xA(x_0+x(\tau))},
\end{equation}
and its \emph{expectation value},
\begin{equation}
  \label{eq:wew1}
  \langle W_{x_0}\rangle
:=
  {\int_{ {x(0)=x(T)\atop\mathrm{CM}}}  \mathcal Dx(\tau)\ 
    W_{x_0}[x(\tau)] \,  e^{-\int_0^Td\tau\ {\dot x^2\over4}}
  },
\end{equation}
we can write:
\begin{equation}
  \label{eq:gammaur}
  \Gamma_\mathrm{E}^1[A]=-{1\over(4\pi)^{D/2}}\int d^Dx_0\int_0^\infty{dT\over
T^{1+{D/2}}}\ e^{-m^2T}
  \langle W_{x_0}\rangle +\mathrm{c.t.}
\end{equation}
Here we have added counterterms (c.t.) which have to be fixed by
renormalization of physical parameters.  If the electric field is nonzero,
$\Gamma^1[A]$ obtains an imaginary part, arising from poles of the Wilson-loop
expectation value $\langle W_{x_0}\rangle$ on the real $T$ axis. Surrounding
the poles by halfcircles in the upper half plain in agreement with causality
leads to
\begin{equation}
    \mathrm{Im}\Gamma_\mathrm{E}^1[A]=-{1\over\left(4\pi\right)^{D/2}}
      \int d^Dx_0\ \mathrm{Im}\sum_{T_\mathrm{pol}}
      {1\over {T_\mathrm{pol}}^{1+{D/2}}}e^{-m^2 T_\mathrm{pol}}
      (-\pi i)\mathrm{Res}\left(\langle W_{x_0}\rangle,
      T_\mathrm{pol}\right),
\label{eq:polsumme}
\end{equation}
where the sum goes over all poles with positions $T_\mathrm{pol}$.
The exponential factor with Gau\ss ian velocity weight in the path
integral in Eq. (\ref{eq:wew1}) suppresses the contribution of long
paths. Therefore, the integral is dominated by paths that tightly
wiggle around the common center of mass. This gives rise to the picture
of a \emph{loop cloud} sitting at $x_0$ and scanning the background
field in the neighborhood of $x_0$.  Hence, the nonlocal nature of the
phenomenon is already apparent in the formalism.

Let us mention in passing that the path integral for a constant $E$
background is Gau\ss ian, can thus be done exactly, and results in
$\langle W_{x_0}\rangle=eET/\sin (eET)$; see below. Summing over the
pole positions of the inverse sine results in the famous Schwinger
formula (for scalar QED in this case),
\begin{equation}
    \mathrm{Im}\Gamma_\mathrm{M}^1[E=\text{const.}]= - \frac{V}{16\pi^3} \,
(eE)^2
\sum_{n=1}^\infty \frac{(-1)^n}{n^2} \, e^{-\frac{m^2}{eE}\, \pi n},
\label{eq:SchF}
\end{equation}
displaying the nonperturbative dependence on $eE$
\cite{Schwinger:1951nm}; here, $V$ denotes the space-time volume. Each
term in the sum corresponds to production of $n$ coherent pairs.

\section{Worldline numerics}

\subsection{Worldline discretization}

The worldline numerical algorithm for the present problem
partly resembles closely those developed in detail in 
\cite{Gies:2001zp,Gies:2001tj,Langfeld:2002vy,Gies:2003cv,Moyaerts:2003ts},
the essential steps of which we will recall in the following for
completeness. As a first step, we introduce the \emph{unit loop} $y(t)$,
\begin{equation}
  y(t):={1\over\sqrt T}x(Tt).
\end{equation}
The Wilson-loop expectation value then reads
\begin{equation}
  \langle W_{x_0}\rangle:={\int_{
  {y(0)=y(1)\atop\mathrm{CM}}}
  \mathcal Dy(t)\ W_{x_0}\, e^{-\int_0^1dt\ {\dot y^2\over4}}
  }.
  \label{expv}
\end{equation}
The exponential velocity distribution is now independent of $T$,
whereas the Wilson loop yields
\begin{equation}
    W_{x_0}=e^{-i e \int_0^1 dt\ \sqrt{T} \dot{y} A\left(\sqrt{T} y +
x_0\right)}.
    \label{wfull}
\end{equation}
For a finite ensemble of paths that are distributed according to the
weight factor $\exp(-\int_0^1 d\tau\ {\dot{y}^2\over4})$,
the Wilson-loop expectation value is equal to the arithmetic ensemble average
of $W_{x_0}$. Since the weight factor for unit loops is independent of $T$ and
$x_0$, such a loop ensemble has to be generated only once for computing
$\langle W_{x_0}\rangle$ for different $T$ and $x_0$.

For the numerics, we discretize the propertime such that each loop $y(t)$ is
represented by a finite number of $N$ points per loop (ppl) $y_k$ at $t=k/N$,
with $k=1,\cdots,N$.\footnote{The worldline points $y_k$ live in continuous
  space-time, $y_k\in \mathbbm{R}^D$. For an alternative lattice formulation,
  see \cite{Schmidt:2002mt,Schmidt:2003bf}.} We use the \emph{vloop} algorithm
\cite{Gies:2003cv} to create an ensemble of $n_\mathrm L$ discrete and closed
unit loops $\{y_k\}$ with the distribution functional
\begin{equation}
P[\{y_k\}]=\delta(y_1+\cdots+y_N)\exp\left(-{N\over4}\sum_{k=1}^N(y_k-y_{k-1}
)^2\right)\label{eq:a1},
\end{equation}
with the condition $y_0\equiv y_N$ for closed loops. Eq.~\eqref{eq:a1}
represents the discrete form of the weight factor $\exp(-\int_0^1 d\tau\ 
{\dot{y}^2\over4})$, with the delta function reflecting the
center-of-mass condition.

For a gauge-invariant discretization of the Wilson loop, the gauge
field, in principle, should be treated as a link variable, i.e., $dt\,
\dot y(t)\, A(y(t)) \to (y_k -y_{k-1})\, A( (y_k + y_{k+1})/2)$, with
the gauge field evaluated at the center of the links. However, the
link centers do not carry the same information about the distribution
of the worldlines in spacetime as the sites $y_k$ do: for instance,
the link centers have a smaller average distance to the center of mass
$x_0$ than the sites do. The use of the link centers actually corresponds
to effectively shrinking the loop cloud. Of course, this difference
becomes irrelevant in the propertime continuum limit $N\to \infty$.
However, for small $N$, this effect leads to sizeable systematic
deviations from the continuum limit. We avoid this systematics by
evaluating the gauge field at the sites instead,
\begin{equation}
  \int_0^1dt\, \dot y\, A(\sqrt T y+x_0)\to \sum_{k=1}^N (y_{k+1}-y_k)\,
  A(\sqrt T  y_k+x_0).
\end{equation}
It turns out that possible violations of gauge invariance for smooth gauges
such as the Lorenz gauge remain much smaller than other systematic and
statistical errors for the background fields studied in this
work.\footnote{In the general case, we, of course, recommend the
  gauge-invariant link variable discretization. In order to reduce the
  systematic error mentioned above, order $1/N$ improvements of the action may
  be useful.}

For the effective action and the pair-production rate, the $T$
integration in Eq. (\ref{eq:gammaur}) has to be performed. For the
simple case of a constant field, this can be done elegantly by a
fast-Fourier transform (FFT) after the $T$ integration has been rotated
onto the imaginary axis. Thereby, the pair production is obtained for
a whole spectrum of masses and field strengths, respectively, all at
once. This procedure and its limitations will be discussed in the
Appendix. However, for more general field configurations, an overlap
problem arises: when performing the $T$ integration, one faces
situations in which the path integral is dominated by very elongated
loops, despite the exponential suppression by the weight
factor. Physically, those virtual pairs that delocalize strongly gain
more energy and have a larger probability of becoming real. In this
case, the finite loop ensemble with only a few elongated worldlines is
no longer representative for the over-countably many paths of the path
integral. To solve this problem, we have developed the routine
presented in the following.  Its cornerstone is a probability
distribution analysis of particular worldline-ensemble properties
along the lines suggested in
\cite{GiesSanchez-GuillenVazquez_in_preparation}.

\subsection{CDF fit for pair production}

In order to motivate our algorithm, let us first consider the case of a
constant homogeneous electric field $E$ in Minkowski space.\footnote{$E$
  denotes the \emph{Minkowskian}, i.e., \emph{physical}, field strength.} This
is related to the Euclidean gauge potential by
$A|_{\text{E}}=(0,0,0,-iEx_1)^\top$. The corresponding Wilson loop can be
written as
\begin{equation}
  W(I)=e^{-TeEI},\quad \text{where}\,\,\,I:=\int_0^1 dt\ \dot{y}_4y_1.
\label{eq:WvonI}
\end{equation}
The scalar quantity $I$ contains all relevant information about the unit loop
for the present case. The probability density function (PDF) of $I$ for our
loop ensembles is defined by
\begin{equation}
P(I)=\int_{y(0)=y(1)\atop\mathrm{CM}}
  \mathcal Dy\, \delta\left(I-\int_0^1 dt\, \dot{y_4} y_1\right)\,
  e^{-\int_0^1dt\, {\dot y^2\over4}}. 
\end{equation}
With the aid of a Fourier representation of the $\delta$ function, the
path integral becomes Gau\ss ian and yields  
\begin{equation}
P(I)={\pi\over4}\cosh^{-2}\left({\pi\over2}I\right)\label{eq:PDFconst}
\end{equation}
for constant fields. In terms of the PDF, the Wilson-loop expectation
value can be written as
\begin{equation}
\langle W\rangle=\int_{-\infty}^\infty dI\
P(I)W(I),\label{eq:Wint} 
\end{equation}
resulting in $\langle W \rangle={TeE/\sin(TeE)}$ in agreement with the
Schwinger pair-production rate for constant fields, cf. Eq~\eqref{eq:SchF}.

For inhomogeneous field configurations, $\langle W_{x_0}\rangle$ can be
computed in a similar way. Generalizing the definition of $I$, 
\begin{equation}
I_{x_0}:={i\int_0^1dt\ \dot yA(\sqrt Ty+x_0)\over \sqrt TE_0}, \label{eq:Iofx}
\end{equation}
the PDF becomes space-time and proper-time dependent, 
\begin{equation}
P_{x_0}(I)=\int_{y(0)=y(1)\atop\mathrm{CM}}
  \mathcal Dy\, \delta\left(I-I_{x_0}\right)\,
  e^{-\int_0^1dt\, {\dot y^2\over4}}.
\end{equation}
But for each space-time point, the Wilson-loop average can still be
computed analogously to Eq.~\eqref{eq:Wint}, with $ W(I) = e^{-T
  eE_o I}$ similar to Eq.~\eqref{eq:WvonI}. The reference field
strength $E_0$ is a priori arbitrary and has been introduced to obtain
a dimensionless quantity.  In most cases, we may use the local field
strength $E_0:=|E(x_0)|$, or some averaged value. For the constant $E$
field, our generalized definition of $I_{x_0}$ conforms to the
previous one. The PDF of $I_{x_0}$ is generally not known analytically
but will be computed numerically from a finite loop ensemble.
Nevertheless, analytical knowledge about $P_{x_0}(I)$ is required,
owing to the following reasons:

\begin{itemize}
  
\item The use of a Monte Carlo algorithm does not only demand the
  worldline spacetime metric to be Euclidean, but also requires the
  contour of the propertime integral to run along the real $T$ axis.
  However, as is already obvious for the constant-field case, the
  integral in Eq. (\ref{eq:Wint}) is well defined only for
  $|TeE|<\pi$. At $|TeE|=\pi$, the first pole $T_{\text{pol}}$ of
  $\langle W_{x_0} \rangle$ is hit. For larger values of $|eET|$, the
  $I$ integral has to be replaced by its analytic continuation, which
  can only be constructed if $P_{x_0}(I)$ is known analytically.
  
\item By using \emph{finite} loop ensembles, we already face an overlap
  problem for small $T$ values: the majority of loops have a
  small $I$ value, whereas those few loops with large $I$ dominate the
  $I$ integral in Eq.  (\ref{eq:Wint}); see the Appendix. A controlled
  extrapolation of the PDF to large $I$ values from reasonably big
  worldline ensembles can thus reduce the numerical cost considerably.
  This can be achieved by fitting the numerical PDF data to an
  analytical ansatz.

\end{itemize}

\noindent
The last point is, of course, related to the nonlocal features of
pair production. The quantity $I$ on the one hand is connected to the
electrostatic energy gain of a virtual pair that propagates in a
background field, and on the other hand roughly measures the space-time
extent of a worldline. The dominance of large $I$ values in the final
result arises from strongly delocalized virtual pairs.

To obtain an analytical estimate for the PDF, we generalize the result
for the constant field, Eq. (\ref{eq:PDFconst}), by the following
ansatz governed by two parameters $\alpha$ and $\nu$:
\begin{equation}
P_{x_0}(I)=N\cosh^{-2\nu}\left({\pi\over2}\alpha I\right).\label{eq:Pfit}
\end{equation}
The parameters control the two main features of the distribution:
width and sheerness. Both parameters depend on the spacetime point
$x_0$ and on the propertime parameter $T$. The normalization constant
$N$ is a function of $\alpha$ and $\nu$ fixed by $\int dI\,
P_{x_0}(I)=1$.  Numerically more convenient is the corresponding
cumulative density function (CDF) of $|I|$,
\[D_{x_0}(|I|)=\int_{-|I|}^{|I|}d\hat I\ P(\hat I).\]
For given values $T$ and $x_0$, we determine $\alpha$ and $\nu$ by a
fit of the numerical data to this CDF. Inserting the resulting
parameters into Eq.  (\ref{eq:Pfit}) yields the desired analytical
expression for $P_{x_0}(I)$.  Performing the $I$ integral in Eq.
(\ref{eq:Wint}) gives $\langle W_{x_0}\rangle$ as function of $\alpha$
and $\nu$,
\[\langle W_{x_0}\rangle=N{4^\nu\over \pi
  \alpha}{\Gamma(\nu+{TeE_0\over\pi \alpha})\Gamma(\nu-{TeE_0\over\pi
    \alpha}) \over\Gamma(2\nu)}.\]
This result also represents the desired analytical continuation to
arbitrary values of $T$ or $|eE_0T|$, and solves the problem of Wick
rotating the result of the Euclidean path integral back to Minkowski
space. We observe that the second Gamma function in the numerator is
responsible for the pole structure of $\langle W_{x_0}\rangle$ on the
positive real axis. Poles occur if
\begin{equation}
\nu-{TeE_0\over\pi \alpha}=-l,\quad \text{with}\,\, l=0,1,2,\cdots.
\label{eq:polbed}
\end{equation}
Since $\alpha$ and $\nu$ depend on $T$, Eq.~(\ref{eq:polbed})
determines the pole positions $T_\mathrm{pol}$ only implicitly; in
practice, we solve for $T_{\text{pol}}$ iteratively. At the pole
location, the corresponding residue is
\[\mathrm{Res}\left(\langle W_{x_0}\rangle,T_{\text{pol}}\right)
=\left.N{4^\nu\over \pi \alpha}{\Gamma(2\nu+l)
\over\Gamma(2\nu)}{(-1)^l
\over l!{d\over dT}\left(\nu-{TeE_0\over\pi \alpha}
\right)}\right|_{T_\mathrm{pol}},\]
which we plug into Eq. (\ref{eq:polsumme}) to obtain the
pair-production rate. For a reliable control of the statistical error,
we perform a jackknife analysis for all secondary quantities. For the
systematic error due to the propertime discretization, we increase the
number of $N$ ppl to approach the continuum limit at least within the
statistical errors.

Obviously, the reliability of our results depends crucially on the
ansatz (\ref{eq:Pfit}) for the PDF. Apart from our consistency
arguments referring to the shape of the PDF and the resulting pole
structure, final support can only be given by nontrivial tests
described below. In summary, the sufficiency of the ansatz is
confirmed by the following arguments:

\begin{itemize}

\item The free parameters control the two most essential features of the
  distribution, the width and the sheerness, which encode, in particular, 
  the important contributions from the strongly delocalized virtual pairs. 
\item The exact functional form for the constant-field limit is
  supported by the ansatz. Even without further checks, we could thus
  expect satisfactory results at least for slowly varying fields.
\item As a nontrivial analytical confirmation, we stress that the ansatz leads
  to a reasonable pole structure of $\langle W\rangle$ that can encode
  information about coherent $n$ pair production. 
\item The ansatz provides highly convincing results for the Sauter
  potential including the constant field limit as special case, as
  presented in the next section. Any systematic deviations from the exact
  result are negligibly small compared to the statistical error.

\end{itemize}

\section{Sauter potential}

The Sauter potential defines an electric field with solitonic profile in one
spatial direction which is constant in all other directions including time. The
direction of the field vector is constant and coincides with the
solitonic-profile direction. An analytical expression of the corresponding
total pair-production rate has been found by Nikishov \cite{Nikishov:1970br}.
In Minkowski space, the Sauter potential reads
\[A^0|_\mathrm M=-a\tanh(kx^1),\ \ A^i|_\mathrm M=0,\quad
E^1|_\mathrm M = \frac{a k}{\cosh^2 (kx^1)}.
\]
The parameter $k$ defines the inverse width of the electric field, whereas $a$
governs its maximum, $E_\mathrm{max}\equiv ak$. The constant-field limit is
recovered for $k\to 0$ for fixed $ak$.

\begin{figure}[h]
\centering
    \includegraphics[width=.5\textwidth]{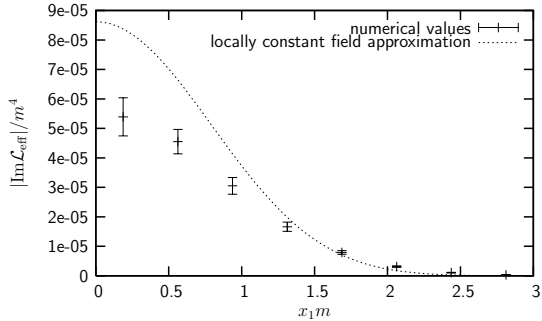}
\caption{Spatial distribution of the effective Lagrangian's imaginary part
  for a Sauter potential. The numerical result is compared to the
  locally constant-field approximation that overestimates the true result by
  up to $\sim50\%$. Parameters of the Sauter potential: $k=0.4m$,
  $E_\mathrm{max}=(m^2/e)$. Parameters of the loop cloud: $n_\mathrm
  L=100000$, $N=1000$ ppl.}
\label{fig:gammasauter}
\end{figure}

As an example, Fig.~\ref{fig:gammasauter} shows the $x_1$ dependence of the
local pair-production rate $\sim\mathrm{Im}\mathcal L_\mathrm{eff}$ for
$k=0.4m$ and $E_\mathrm{max}\equiv ak=(m^2/e)$ computed by our algorithm.
It is compared to the approximated effective Lagrangian obtained by a
derivative expansion to lowest order, i.e., by assuming the field to be
locally constant (Schwinger formula).  We observe that the local rate
predicted by the algorithm is spatially smeared compared to the Schwinger
formula.

\begin{figure}[t]
\centering
    \includegraphics[width=.5\textwidth]{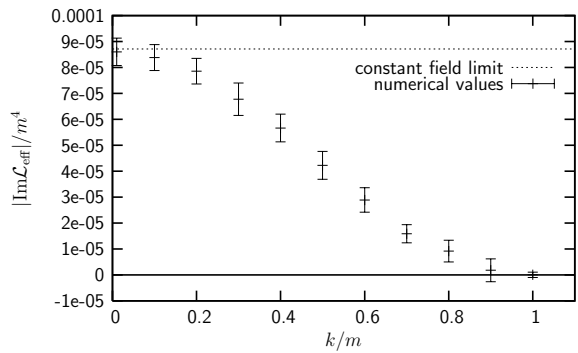}
\caption{The imaginary part of the effective Lagrangian in the center
  of a Sauter potential with maximal field strength $m^2/e$
  versus the inverse width parameter $k$.  The dashed line marks the
  analytically obtained contribution of the first pole to
  $|\mathrm{Im}\mathcal L_\mathrm{eff}|$ for the constant-field limit
  (higher poles give a 1\%\ correction). $n_\mathrm L=100000$,
  $N=1000$ ppl.}
\label{fig:centresauter}
\end{figure}

The pair-production density in the center $x^1=0$ of the Sauter potential with
maximal field strength $E_\mathrm{max}=(m^2/e)$ is shown in Fig.
\ref{fig:centresauter} versus the width parameter $k$; units are set by the
electron mass scale. For large width $k\to 0$, the constant-limit is
approached, and our CDF fit algorithm correctly reproduces the Schwinger
formula. The more interesting limit occurs for $k=m$ where the production rate
vanishes. Even though the electric field is still nonzero, the width of the
Sauter potential is equal to the Compton wavelength. Therefore, even if a
virtual pair delocalizes completely along the direction of field lines with
the $e^{-}$ going to $x^1\to \infty$ and the $e^{+}$ going to $-\infty$, the
pair cannot acquire enough energy to become real. This important physical
example is missed completely by the locally constant-field approximation,
emphasizing the role of nonlocalities. 

Moreover, the limiting case of $k\to m$ is an extreme and crucial test
for our algorithm based on the PDF ansatz \eqref{eq:Pfit}: in the
vicinity of this limit, there is literally not a single worldline in
our finite ensemble that exhibits the strong delocalization required
for giving rise to a {\em direct} contribution to the final result
(the number of sufficiently elongated worldlines is exponentially
suppressed). Nevertheless, the overall distribution of $I$ values
allows for a controlled extrapolation via the CDF fit, leading to a
numerical estimate even for the directly inaccessible regime. As a
measure for the resulting error, we mention that our result for the
case $k=m$ is not exactly zero, but $|\text{Im}
\mathcal{L}_{\text{eff}}|/m^4 =5.73 \cdot 10^{-8} \pm 1.03\cdot
10^{-6}$. 
We conclude that possible systematic errors induced by our
CDF fit algorithm are negligibly small compared to the statistical
error.

\begin{figure}[h]
\centering
    \includegraphics[width=.5\textwidth]{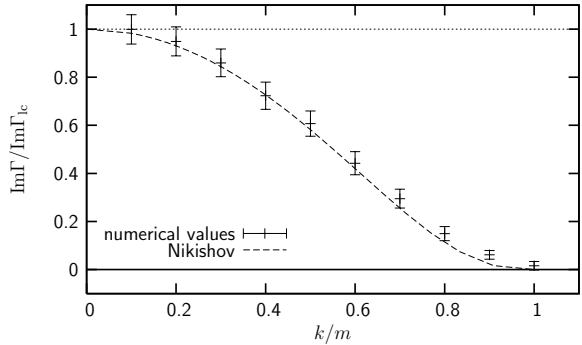}
\caption{The imaginary part of the effective action for a Sauter
  potential as fraction of the locally constant-field approximation
  $\mathrm{Im}\Gamma_\mathrm{lc}$ versus the width parameter $k$ in units of
  $m$: comparison of the numerical result with Nikishov's analytic expression.
  $n_\mathrm L=100000$, $N=1000$ ppl.}
\label{fig:gammasauterintvgl}
\end{figure}

Finally, Fig. \ref{fig:gammasauterintvgl} shows the integrated total
pair-production rate $\mathrm{Im}\Gamma$ compared to
the Nikishov result. The agreement is satisfactory and the vanishing pair
production for $ea=m$ is reproduced within the error bars.

\section{Sine-modulated potential}
\label{SecSine}
In this section, we study the superposition of a spatially varying
sine potential with a constant field. This configuration is of general
interest, as it is representative for a class of field configurations
which are superpositions of a slowly varying field---in our example
the constant field---and higher-oscillation modes.  A very important
aspect is the dependence of the pair-production rate on the spatial
oscillation frequency of the small-scale field structures.  We
consider this example as a paradigm for the role of nonlocal phenomena
in pair production.

In Minkowski space, the potential is given by
\[A^0|_\mathrm M=-a\sin(kx^1)-E_0x^1,\ \ A^i|_\mathrm M=0.\ \]
It corresponds to an $E$ field in $x^1$ direction with field strength
\[E^1|_\mathrm M=E_0+ak\cos (kx^1),\]
which has extremal field strength of $E_\mathrm{max,min}=E_0\pm ak$.
As an example, we study a field with $E_0=0.2(m^2/e)$ and
$E_\mathrm{max}=0.3(m^2/e)$.

\begin{figure}[h]
\centering
    \includegraphics[width=.5\textwidth]{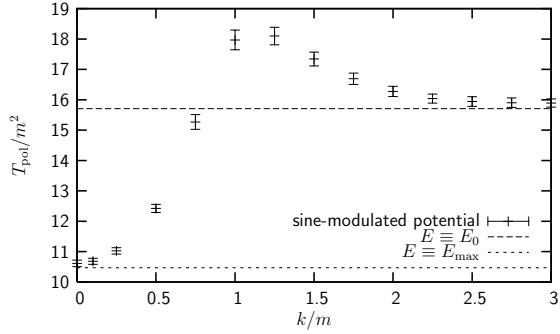}
\caption{Position of the first pole $T_{\text{pol}}$ of $\langle W\rangle$ on
  the real proper-time axis at a maximum of the field strength. With increasing
  frequency $k$, the pole moves from the constant-field limit $E\equiv
  E_\mathrm{max}$ to the limit $E\equiv E_0$.  Parameters of the field:
  $E_0=0.2(m^2/e)$, $E_\mathrm{max}=0.3(m^2/e)$.  In between, it
  develops an unexpected maximum corresponding to a minimum of the local
  production rate.  Parameters of the loop cloud: $n_\mathrm L=100000$,
  $N=1000$ ppl.}
\label{fig:ppsin}
\end{figure}

Figure \ref{fig:ppsin} shows the position of the first pole $T_{\text{pol}}$
of the Wilson-loop expectation value on the real propertime axis for $x^1$ in
the center of a maximum of the field strength.  For small $k$, the pole
position of the constant-field limit $E\equiv E_\mathrm{max}$ is reproduced.
For large $k$, the pole position converges to the result of the averaged field
$E\equiv E_0$.  In between, the curve is not monotonically increasing, as one
might have expected, but reaches $T$ values which are significantly larger
than in both limiting cases.  As a consequence, the corresponding
local production rate will be smaller than in
the constant-field limit $E\equiv E_0$.

This behavior is, of course, a consequence of nonlocalities and can be easily
understood in the worldline picture in terms of loop clouds:
\begin{figure}[h]
\centering
    \includegraphics[width=.5\textwidth]{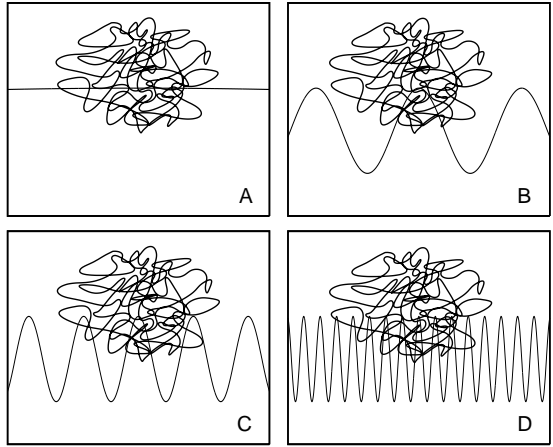}
    \caption{An artist's view on a loop cloud (worldline ensemble) at a
      maximum of the field strength. For small frequencies, it detects only
      the maximum (a).  After increasing the frequency, the two nearest minima
      dominate (b).  For larger frequencies the cloud encounters further
      maxima (c), until it perceives an averaged field (d).  }
    \label{fig:sinserie}
\end{figure}
Starting with the limit $k\to0$, a loop cloud sitting at a maximum
detects a constant field of strength $E_\mathrm{max}$.  A sketch of
this scenario if given in Fig. \ref{fig:sinserie}(a).  If $k$ is
increased and the wavelength of the sine becomes shorter, the loop
cloud overlaps more and more with the minima on either side of the
maximum and the pole moves to larger $T$ values.  If $k$ exceeds a
certain value, in our example at about $k=0.8m$, the two minima close
by dominate the Wilson-loop expectation value, Fig.
\ref{fig:sinserie}(b). Despite the maximum in the center of the loop
cloud, the pole is at a larger $T$ value than for the averaged field.
Not until the loop cloud approaches the adjacent maxima, Fig.
\ref{fig:sinserie}(c), do the $T$ values become smaller again, finally
converging to the value of the averaged field, Fig.
\ref{fig:sinserie}(d).

Since the Wilson-loop expectation value at a maximum of the field
strength can be dominated by the adjacent minima, the inverse
situation can also occur at a minimum where the result can be dominated by
the two adjacent maxima. In this case, the first pole of $\langle
W\rangle$ is at a \emph{smaller} $T$ value than for the averaged
field, leading to a \emph{larger} imaginary part of the effective
Lagrangian.
\begin{figure}[h]
\centering
    \includegraphics[width=.5\textwidth]{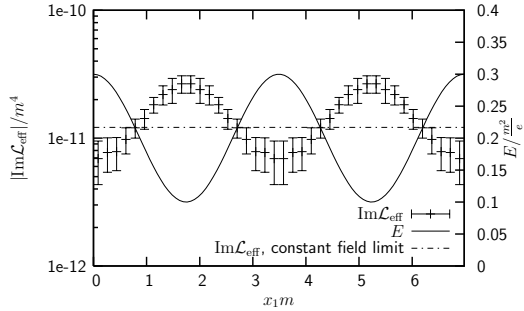}
\caption{Spatial distribution of the imaginary part of the effective-action
  density for the sine-modulated potential with $k=1.8m$ compared to
  the constant-field limit $E\equiv E_0$. Nonlocal effects lead to the
  seemingly paradoxical phenomenon that the pair-production rate is
  maximal at the field-strength minima and vice versa.
  $n_\mathrm{L}=200000$, $N=1000$ ppl.}
\label{fig:gammasin}
\end{figure}
This inversion is shown in Fig. \ref{fig:gammasin}, where the spatial
distribution of the imaginary part of the effective action for
$k=1.8m$ is plotted in comparison to the constant-field limit $E\equiv
E_0$. We observe that the nonlocalities induce a seemingly paradoxical
phenomenon in this case: the maxima of the local pair-production rate
occur at the minima of the electric field strength and vice versa.
\begin{figure}[h]
\centering
    \includegraphics[width=.5\textwidth]{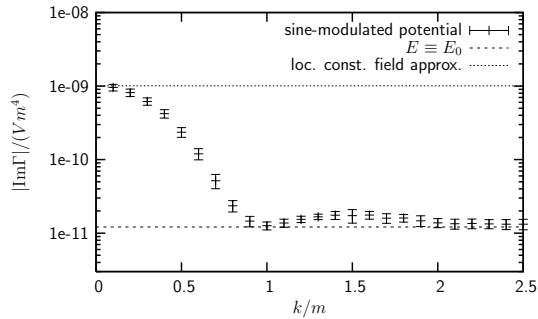}
\caption{The imaginary part of the total effective action
  per space-time volume against the frequency $k$. The dashed lines
  mark the locally constant-field approximation and the result for the
  averaged field $E\equiv E_0$, respectively. The former (dashed
  lines) misses the true result by an order of magnitude already for
  $k/m\simeq 0.5$.  $n_\mathrm{L}=200000$, $N=1000$ ppl. }
\label{fig:gammasinint}
\end{figure}

Figure \ref{fig:gammasinint} depicts the imaginary part of the total
effective action per space-time volume for our example configuration
versus the frequency $k$.  In contrast to its density at $x_0$,
$\mathrm{Im}\Gamma$ does not fall below the result for the averaged
field. For oscillation frequencies near $k=0$, we observe that the
locally constant-field approximation based on the derivative expansion
fails rather early by an order of magnitude for $k\simeq0.5 m$; this
is remarkable, since the effective expansion parameter
$k^2/m^2\simeq0.25$ might have been considered as small enough.

In the opposite limit, for large
 frequencies $k$, we obtain the averaged
 constant-field limit $E\equiv E_0$. It is
remarkable that the imaginary part of the effective action reaches the
value of the averaged field for $k$ values as small as about $k=m$,
whereas its density still fluctuates spatially for even larger $k$
values, as seen in Fig.  \ref{fig:gammasin}. The fluctuations cancel
each other, so that they have no effect on the integrated quantity.
The numerical accuracy does not eliminate the possibility of a
$k$-dependent structure for $k$ values larger than $m$.  According to
the values of Fig. \ref{fig:gammasinint}, the central values suggest a
slight increase of the pair production for $k>m$, until it falls back
to the result for the averaged field if $k/m\to\infty$.  To definitely
clarify this, larger loop ensembles are necessary at the expense of
CPU time.  However, the present result shows that any possible $k$
dependence for $k>m$ has to be relatively small and the averaged-field
approximation yields good results in this range. 

Let us finally compare our results for the spatially sine-modulated
field with those for spatially homogeneous fields with time
dependencies. Especially the case of an electric field oscillating in
time with frequency $\omega$ has been studied with WKB methods
\cite{Brezin:1970xf,Popov:1972,Popov:1973az,Piazza:2004sv} which
were originally developed for ionization processes in atomic physics
\cite{Keldysh:1965}. The nature of pair production in this case
depends on the size of the ``adiabaticity parameter''
$\gamma:=m\omega/(eE)$; for small $\gamma\ll 1$, the result approaches
the Schwinger formula and pair production thus is a nonperturbative
phenomenon. For large $\gamma\gg 1$, the result becomes perturbative
in $(eE)/(m\omega)$ and pair production arises from multiphoton
scattering. In our case, we can, of course, also form a similar
parameter\footnote{This parameter is not unique in our case, since the
parameter $a$ of the sine modulation introduces yet another scale.}
$\tilde \gamma=mk/eE_0$, with $\tilde \gamma$ small or large roughly
corresponding to the two limiting cases discussed above. However, it
is important to stress that pair production is nonperturbative in both
limits for our sine-modulated field. In particular, the large-$k/m$
(or large $\tilde\gamma$) case cannot be understood in terms of
multiphoton processes. Taking the external field to all orders into
account is essential for the final result.

\section{Conclusion}

We have developed a new universal approach for computing local
production rates for spontaneous pair creation by the Schwinger
mechanism in scalar QED. Our method is based on the combination of the
worldline formalism with Monte Carlo techniques. As a first result, we
have not only rediscovered Nikishov's analytic result for the total
pair-production rate in a Sauter potential, but moreover we have
computed the local pair-production rate for this classic case for the
first time. Most importantly, the algorithm is not restricted to any
spatial symmetry of the given background potential but is applicable
for arbitrary potentials.

As a nontrivial example, we have applied the algorithm to a constant
electric field modulated by a spatial sine oscillation. This field
configuration is representative for a whole class of fields with
large-scale structures and small-scale oscillations. By varying the
spatial oscillation frequency, qualitatively different features of
pair production can be investigated. For small frequencies, our
numerical result agrees with the derivative expansion to lowest order;
the latter breaks down completely for spatial variations on the order
of a few times the Compton wavelength. On this length scale and below,
our results show clearly that another approximation scheme becomes
reliable: the local production rate can well be approximated by
inserting the {\em spatially averaged} field into the Schwinger
formula. This averaged-field approximation can be trusted on the few-percent
level for spatial variations of the size of the Compton
wavelength. We would like to emphasize that the small validity bound
of the derivative expansion for the imaginary part of the effective
action density is not related to the same observation for the real
part, as discovered in \cite{Langfeld:2002vy}; the latter arises from
a subtle interplay between nonlocal quantum contributions and local
counterterms, whereas the imaginary part is not affected by
renormalization counterterms. Furthermore, the derivative expansion
for the real part of the {\em integrated} effective action works well
even for Compton-scale variations \cite{Graham:2004jb}, whereas it
breaks down early for the imaginary part, as displayed in
Fig.~\ref{fig:gammasinint}.

Apart from these quantitative results for the particular field
configurations considered here, our findings emphasize the crucial
role of nonlocalities in the phenomenon of pair production. Without
the feature of delocalization of a virtual pair, spontaneous vacuum
decay would not occur. The worldline picture underlying our algorithm
is particularly powerful in capturing these nonlocalities and,
moreover, understanding their consequences in an intuitive way.
Especially our results for local pair-production rates illustrate the
nature and the role of nonlocalities transparently. For instance, the
seemingly paradoxical situation that maxima of pair-production rates
can occur at minima of the field strength (cf. Sect.~\ref{SecSine})
cannot be understood from a local approximation. However, the
worldline picture identifies a natural explanation of this phenomenon
in terms of the delocalization properties of the virtual pairs
described by the worldline trajectory.

From a technical perspective, we have developed a numerical
Monte Carlo algorithm that on the one hand requires a Euclidean
formulation for the quantum fluctuations, but on the other hand
produces reliable results for truly Minkowski-valued physical
observables. The inherent overlap problem is solved in the present
context by a physically motivated ansatz for a suitable cumulative
distribution function (CDF) to which the numerical data can be fitted
and that can be analytically continued to Minkowski space. Even though
the success of this procedure depends strongly on the problem at hand,
we believe that such techniques can be useful in other
Minkowski-valued problems as well. 
The algorithmic strategy itself has been invented in the context of
Euclidean field theory
\cite{GiesSanchez-GuillenVazquez_in_preparation}, where it has turned
out to be highly powerful in a study of nonperturbative worldline
dynamics.

Several extensions of our work are desirable and possible. So far, we
have only considered spatial inhomogeneities, but any realistic field
configuration will also exhibit variations in time. In fact, timelike
variations bring in a new complication, since our Monte Carlo
worldlines live in imaginary time, whereas physical fields depend on
real time. Therefore, our algorithm is directly applicable to all
those cases where the physical field is known analytically, such that
its analytic continuation to imaginary time can be evaluated and
plugged into the numerics. For instance, the exact result for a
solitonic profile in time direction as solved in \cite{Dunne:1998ni}
will be a benchmark test for such an investigation.

Furthermore, our results can, in principle, straightforwardly be
generalized to ordinary spinor QED. As a new complication, the Pauli
term $\sim \sigma_{\mu\nu} F_{\mu\nu}$ occurs in the worldline
integrand. Since this term depends also on the worldline trajectory,
the probability distribution function (PDF) of the ensemble will not
only depend on the quantity $I$ as defined in Eq.~\eqref{eq:Iofx}, but
also on the worldline averaged Pauli-term exponential; let us denote
the latter with $J$, which is also a scalar. Our algorithm might be
generalized as follows: first, compute the PDF of $J$ from the
ensemble and bin the loops according to their $J$ value. Then, apply
the present algorithm to each $J$ bin separately; in particular, the
same analytic-continuation technique can be used. Finally, integrate
over $J$ with the aid of the PDF of $J$. It is important to note that
the $J$ integral can be done last, since the Pauli-term worldline
average cannot induce any poles for the proper-time integral. Of
course, since each relevant $J$ bin has to contain sufficiently many
worldlines, this generalization of our algorithm will at least be an
order of magnitude more time consuming than the one for scalar QED.
At this point, we should stress that the computations for the present
work have still been performed on ordinary desktop PC's.

Finally, it is instructive to compare our method to the instanton
technique of \cite{Affleck:1981bm,DunneSchubertInPrep}, where the
instanton approximation of the worldline integral has been shown to
give the leading-order contribution to pair production. For instance,
in the constant-field case, the one-pair-production rate is generated
by one instanton which is a circular loop. Small fluctuations around
this path lead to the correct imaginary prefactor.  In comparison to
this, our worldlines are extraordinarily complex.  Not a single
worldline loop in our ensembles resembles a circle or
fluctuations thereof. This gives rise to the conjecture that the
computation of the imaginary part requires very little information
about the shape of the loops. We expect that we should be able to
extract the instantonic content of our loops by a suitable cooling
procedure that removes large-amplitude fluctuations. In view of the
success of the instanton approximation, only instantonic plus
small-amplitude-fluctuation information appears to be relevant for
pair production. This agrees with our observation that pair production
is induced by delocalized ``large'' loops that can acquire enough
energy in the $E$ field.  Therefore, it is well possible that a
different loop discretization which optimizes instantonic properties
allows for an even more efficient computation of the imaginary part.
A further investigation of this topic may lead to an even deeper
understanding of pair production.

\section*{Acknowledgment}

We are grateful to G.V.~Dunne, J.~H\"ammerling, K.~Langfeld,
J.~Sanchez-Guillen, M.G.~Schmidt, C.~Schubert, I.-O.~Stamatescu, and
R.~Vazquez for many useful discussions, and to G.V.~Dunne for helpful
comments on the manuscript. This work was supported by the Deutsche
Forschungsgemeinschaft (DFG) under contract Gi 328/1-3 (Emmy-Noether
program).

\begin{appendix}

\section{Constant field: straightforward approach}

\begin{figure}[h]
\centering    \includegraphics[width=.5\textwidth]{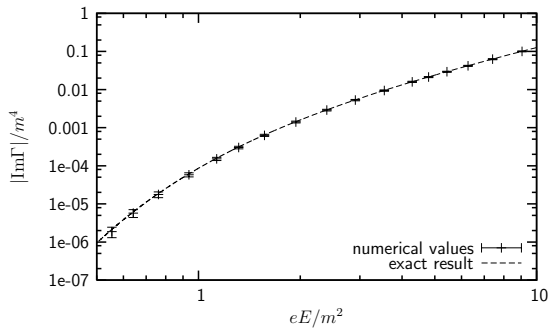}
    \caption{Imaginary part of the effective action obtained by FFT.
$n_\mathrm{L} = 5000$, $N = 1000$ ppl.}
    \label{fig:imgammaw}
\end{figure}

In the following, we present a straightforward realization of
worldline numerics for calculating the pair-production rate in the
constant-field case. The algorithm presented here is an immediate
generalization of the standard algorithm successfully used for the
real part of the effective action
\cite{Gies:2001zp,Gies:2001tj,Langfeld:2002vy,Gies:2003cv,Moyaerts:2003ts}. 
For a constant field in four dimensions, Eq. (\ref{eq:gammaur}) reads
\begin{equation}
  \Gamma_\mathrm{E}^1=-{1\over(4\pi)^2}\int_0^\infty{dT\over T^3}
  e^{-m^2T}\int d^4x_0\left(\langle e^{-TeEI}\rangle-1
  -{1\over6}T^2e^2E^2
  \right),
\end{equation}
where $I$ is defined as in Eq. (\ref{eq:WvonI}), and the counterterms
for on-shell renormalization are included. Rotating the $T$
integration contour onto the imaginary axis and substituting $s=-iTeE$
yields a Fourier integral,
\begin{equation}
  \Gamma_\mathrm{E}^1=\left({eE\over4\pi}\right)^2\int_0^\infty{ds\over s^3}
  e^{-i{m^2\over eE}s}\int d^4x_0\left(\langle e^{-iIs}\rangle-1
  +{1\over6}s^2
  \right). \label{eGamF}
\end{equation}
If the worldline-ensemble average $\langle e^{-iIs}\rangle$ can be
computed reliably, Eq.~\eqref{eGamF} offers a highly efficient
algorithm with the aid of the FFT: in this
case, $\Gamma^1_{\text{E}}$ can be computed for a whole spectrum of
frequencies $m^2/eE$ all at once with FFT. The resulting
imaginary part is shown in Fig. \ref{fig:imgammaw}.
It is highly remarkable that this numerical procedure gives
satisfactory results in a wide range of scales, extending over five
orders of magnitude, with little consumption of CPU time. However, the
algorithm fails for small field strengths. The precise limit is given
by the size of the largest loops in the finite loop ensemble: only
loops with $|I|$ values larger than $m^2/eE$ contribute to the
imaginary part of $\Gamma^1_{\text{E}}$. For weak fields, this implies
that only a few or even no loops contribute and the computation fails.

Beside this problem which is already relevant for the constant-field
case, there is a second limitation. For a different contour in the
complex $T$ plane which supports large $\mathrm{Re}T$ values, the
Wilson-loop expectation value is dominated by the loop with the
largest $I$ value. The Monte Carlo algorithms break down here, since
the error bars become as large as the central value. In general, for
inhomogeneous background fields, it is not possible to find a suitable
integration contour to avoid this problem.

These limitations of the straightforward approach are a manifestation
of the fact that the Euclidean worldline ensemble has insufficient
overlap with Minkowski-valued observables for weak fields.

\end{appendix}

\end{document}